\documentclass[twocolumn,prb,aps]{revtex4}

\usepackage{graphicx}
\usepackage{graphics}
\usepackage{amsmath}
\usepackage{amssymb}
\usepackage{enumerate}

\begin{document}
\title{Coupling of distant parts of pyrrole molecule through virtual and resonant states}

\author{Ragesh Kumar T P$^1$}
\author{P. Nag$^1$}
\author{M. Rankovi\'{c}$^1$}
\author{T. F. M. Luxford$^1$}
\author{J. Ko\v{c}i\v{s}ek$^1$}
\author{Z. Ma\v{s}\'{i}n$^2$}
\author{J. Fedor$^1$}

\affiliation{$^1$J. Heyrovsk\'{y} Institute of Physical Chemistry, Czech Academy of Sciences, Dolej\v{s}kova 3, 18223 Prague 8, Czech Republic}
\affiliation{$^2$Faculty of Mathematics and Physics, Charles University, Institute of Theoretical Physics, V Hole\v{s}ovi\v{c}k\'{a}ch 2, 18000 Prague, Czech Republic}

\begin{abstract}
We experimentally show that the N-H bond cleavage in pyrrole molecule following resonant electron attachment is allowed and controlled by the motion of the atoms which are not dissociating, namely of the  carbon-attached hydrogen atoms. In order to interpret these findings, we have developed a method for locating all resonant and virtual states of an electron-molecule system in the complex plane, based on all-electron R-matrix scattering calculations. Mapping these as a function of molecular geometry allows us to separate two contributing dissociation mechanisms: a $\pi^*$ resonance formation inducing strong bending deformations and a non-resonant $\sigma^*$ mechanism originating in a virtual state. The coupling between the two mechanisms is enabled by the out-of-plane motion of the C-H bonds and we show it must happen on an ultrafast few-fs timescale.

\end{abstract}
\maketitle
The most straightforward way how a molecular bond can be broken is by its direct prolongation without any other significant geometry change. In many cases, such a simple cleavage is symmetry forbidden. If the dissociation is initiated by a vertical transition, e.g, by an electron or photon impact, this type of situation arises when the initial state does not asymptotically correlate with the product states. In order for the bond to be broken, the molecular geometry must distort during the dissociation and thus lift the symmetry constraint. 

This happens, for example, in dissociative electron attachment (DEA) to unsaturated organic molecules and biomolecules. Since DEA is a resonant process allowing for bond dissociation at sub-excitation energies, it attracts a significant interest in the fields of radiation damage to biological tissue~\cite{gorfinkiel_2017} and drug metabolism.~\cite{pshenichnyuk_2018}
It has been postulated for the C-H bond cleavage in acetylene~\cite{orel_acet08} and HCN,\cite{Orel_Chourou_HCN_2011} C-Cl bond cleavage in chlorobenzene~\cite{chlorobenzene} and N-H bond cleavage in nucleobases~\cite{Burrow_VFR_2006} that the coupling between the entrance state (non-dissociative $\pi^*$) and a dissociative state is mediated by an out-of-plane (or out-of-line) motion of the dissociating atom. It has been hypothesized that other molecular parts could be involved in the necessary symmetry lowering~\cite{rescigno_formic06, janeckova_formic13} but such effect has never been experimentally proven. 
Here, we show that in the pyrrole molecule the motion of distant atoms influences the probability of bond cleavage. This effect can have significant consequences for the above mentioned fields.
Pyrrole is a heteroaromatic five-membered ring which represents a common motif in biomolecules.
It is highly symmetrical ($C_{2v}$) but complex which allows us to control and decouple the symmetry-breaking and symmetry-preserving contributions to the dissociation. 

\begin{figure*}[tb]
\includegraphics[width=0.9\textwidth]{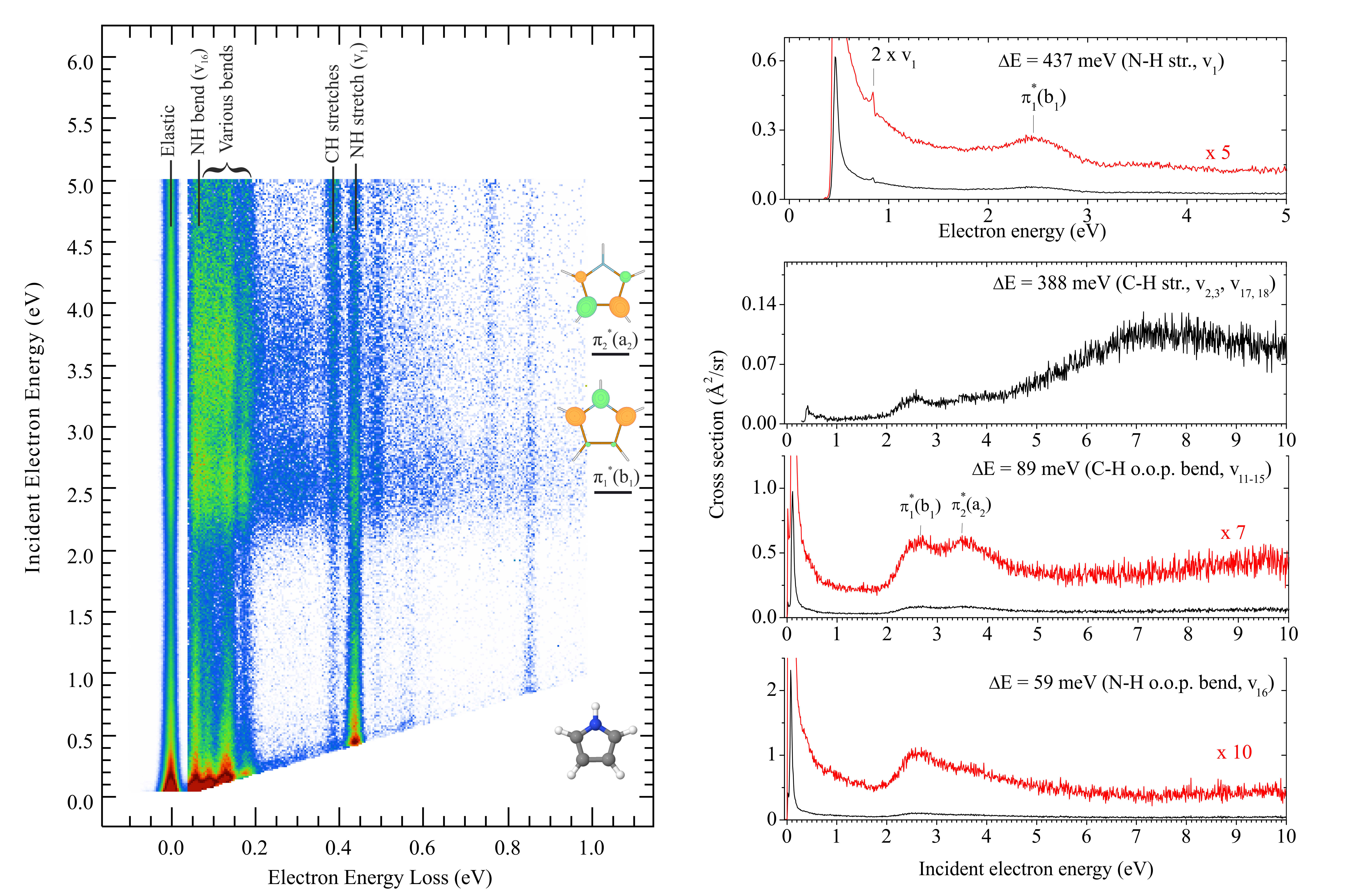}
\caption{Two-dimensional electron energy loss spectrum of pyrrole (left) and the excitation curves of selected vibrations (intensity profiles). The excitation curves were recorded separately in a longer energy range than the 2D spectrum. The N-H stretch vibration (top) is shown in shorter energy range for a better visibility of the marked cusp.}
\label{fig:2D}
\end{figure*}

In order to get the information about the direction of the induced nuclear motion upon resonance formation, we utilized 2D electron-energy loss spectroscopy~\cite{regeta13, anstoter20} using the electrostatic hemishperical spectrometer\cite{Allan_1995, Allan_N2_2005} 
Figure~\ref{fig:2D} shows a 2D electron energy loss spectrum of pyrrole. The electrons collide with a molecule at an incident electron energy $E_i$ (vertical axis) and leave with residual energy $E_r$. Their difference, the energy loss $\Delta E = E_i - E_r$ is plotted on the horizontal axis. The vertical trails thus correspond to the excitation of individual vibrations. Pyrrole has 24 normal modes~\cite{mellouki01}. Not all of them are fully resolved in the EELS experiment (resolution of 18~meV), however, several vibrations or their groups can be unambiguously assigned .  The softest mode is the N-H out-of-plane bend (according to mode numbering from Ref.~\onlinecite{mellouki01}, $v_{16}$, $\Delta E = 59$~meV), the stiffest is the N-H stretch ($v_1$, $\Delta E = 437$~meV). There is a prominent group of C-H out of plane bending vibrations ($v_{11}$, $v_{13}$, $v_{14}$) centered around $\Delta E = 89$~meV) followed by other types of bending (ring deformations and C-H,N-H in plane bents). At 2.5~eV incident energy, the 2D spectrum clearly shows excitation of bending overtones. 

Figure~\ref{fig:2D} also shows the excitation curves of selected vibrations. The IR active modes show threshold peaks which are common for polar molecules.~\cite{itikawa97} 
The threshold peak of the N-H stretch vibration $v_1$ shows sharp cusp at the energy which corresponds to opening of the overtone excitation of this mode. This is evidence of long-range electron-molecule interaction~\cite{Hotop2003, fabrikant16}.
There are also three resonances, centered around 2.45 eV, 3.5 eV and 7.5~eV (very broad). The first two are $\pi_1^* (b_1)$, $\pi_2^* (a_2)$, it will be shown below that the high-energy band is an overlap of $\sigma^{*}(a_{1})$ and $\sigma^{*}(b_{2})$. The two $\pi^*$ resonances were previously detected  by electron transmission experiments~\cite{modelli04, ufa19} and calculations~\cite{oliveira10}, while the two $\sigma^{*}$ resonances have been newly identified using our Siegert approach described below.

\begin{figure}[tb]
\includegraphics[width=0.38\textwidth]{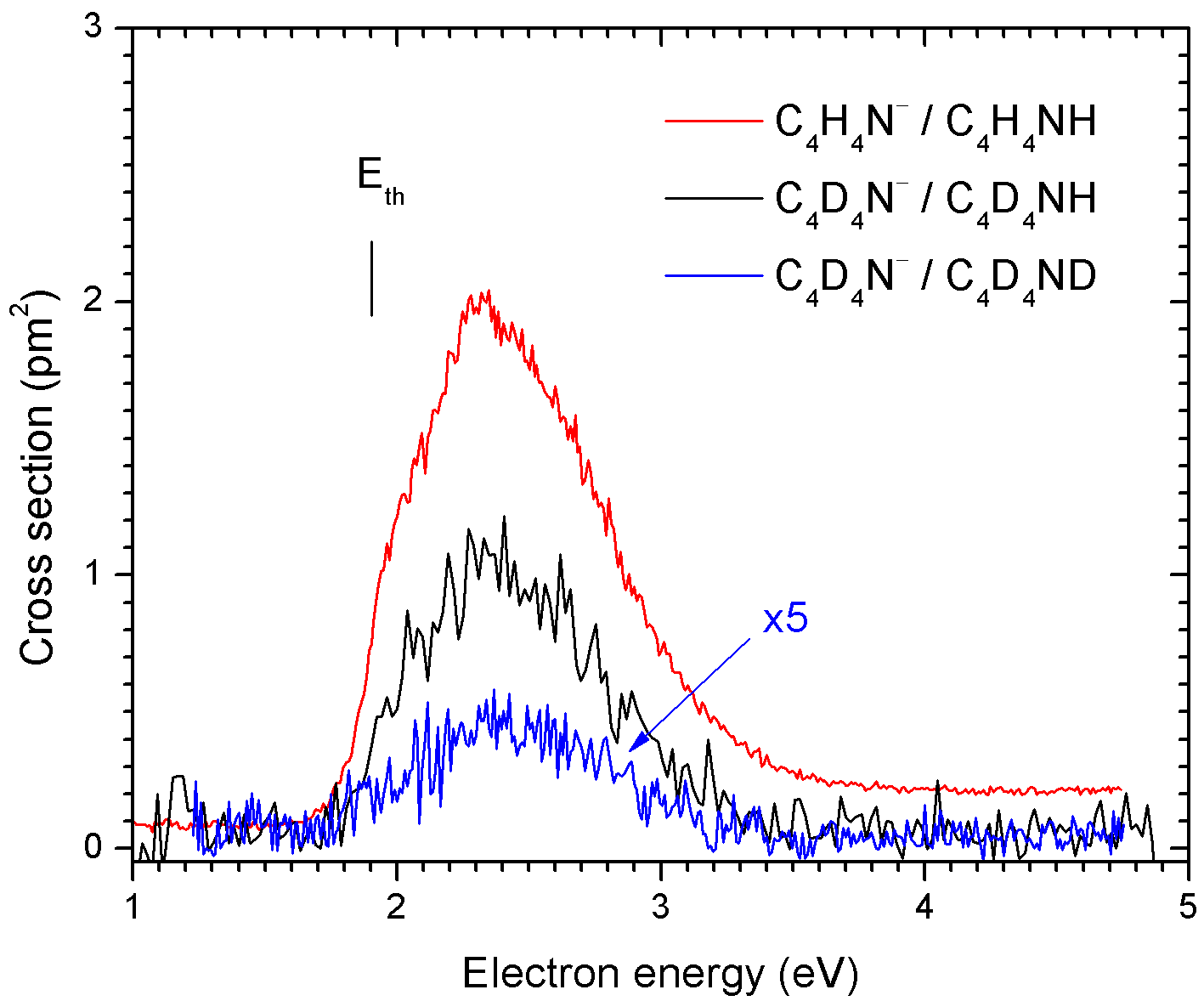}
\caption{Dissociative electron attachment cross section for the N-H bond cleavage in pyrrole (red), fully deuterated pyrrole (blue), and pyrrole partially deuterated on the C-sites (black). The energetic threshold ($E_{th}$) for the process is marked.}
\label{fig:DEA}
\end{figure}

We also measured absolute DEA cross sections for various pyrrole isotopomers using the quantitative DEA spectrometer with a time-of-flight analyzer~\cite{may_acet09, nag19}. The cross section in pyrrole was calibrated against the O$^-$ cross section from CO$_2$ at 4.4~eV (13.3 pm$^2$ eV)~\cite{janeckova_formic13}. The error of the absolute value is estimated to be $\pm$ 20\%. The absolute cross sections of pyrrole isotopomers were then calibrated with respect to each other. Here we did short measurements of ion signal at four different energies across the DEA band and determined the signal ratios. The short acquisition times ensured high stability of electron current and sample pressures. The shape of the DEA band for each isotopomer was then measured on the DEA spectrometer with quadrupole mass filter,~\cite{stepanovic99} which has the electron energy resolution of approximately 70~meV. These spectra were normalized to the time-of-flight absolute values using the invariant of the energy-integrated cross sections.

Figure~\ref{fig:DEA} shows the DEA cross section for the N-H bond cleavage 
\begin{equation}
\mbox{C}_4\mbox{H}_4\mbox{NH} + \mbox{e}^- \to \mbox{C}_4\mbox{H}_4\mbox{N}^- + \mbox{H}.
\label{eq:dea}
\end{equation}
The energy-integrated cross section (quantity independent of the instrumental resolution) is 2.19~pm$^2$ eV. The shape of the band agrees with previous reports of the relative ion yield.~\cite{skalicky04, mufthakov94} 

The DEA cross section is determined by the competition of the dissociation and electron detachment. If one prolongs the timescale for the dissociation, e.g., by replacing an H-atom by deuterium, the dissociative cross section drops, since the electrons have more time to autodetach. 
In pyrrole, the cross section for the fully deuterated compound
is 0.09~pm$^2$ eV, almost a factor of 25 smaller than in reaction~(\ref{eq:dea}). 
Much more interestingly, the reaction    
\begin{equation}
\mbox{C}_4\mbox{D}_4\mbox{NH} + \mbox{e}^- \to \mbox{C}_4\mbox{D}_4\mbox{N}^- + \mbox{H}.
\label{eq:dead4}
\end{equation}
shows a cross section of 0.98~pm$^2$ eV, roughly 50\% of reaction~(\ref{eq:dea}), even though it is still the N-H bond which is dissociated. Such a strong drop cannot be caused by a slight difference in the reduced masses of the dissociating complex (see SI). Instead, the carbon-attached hydrogens have to play a crucial role in the cleavage of the N-H bond since their slowdown significantly reduces the DEA cross section. 

\begin{figure}[tb]
\includegraphics[width=0.5\textwidth]{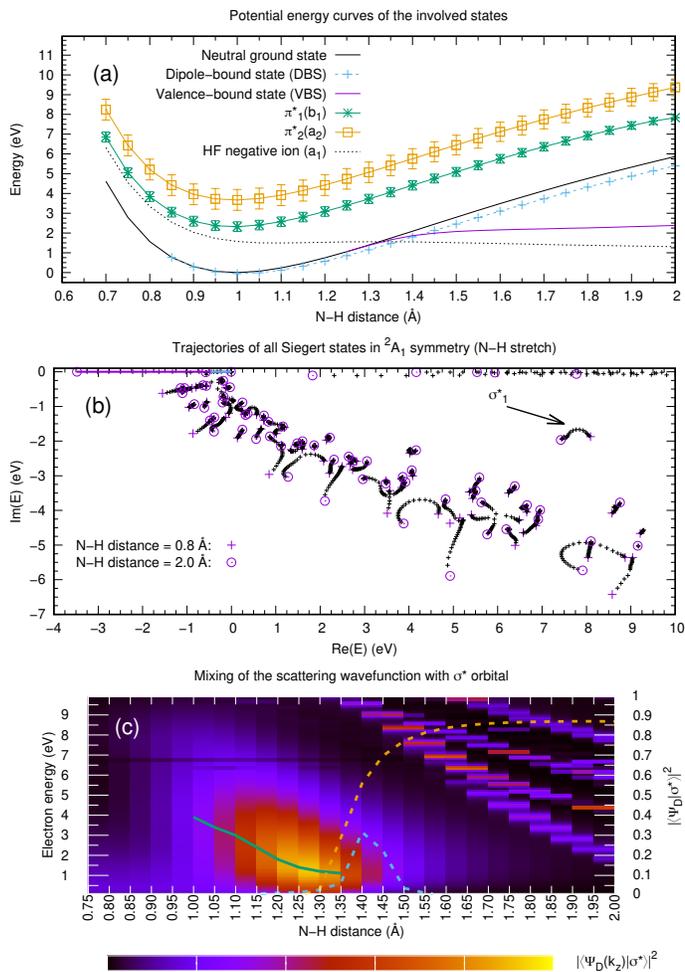}
\caption{\textbf{(a)} Potential energy curves of the neutral, dipole-bound (DBS), valence-bound (VBS) states and the two $\pi^{*}$ resonances.
Resonant energies and widths were obtained directly from the complex plane. The widths are indicated by the 'error bars'. A curve for the lowest $^{2}A_{1}$ anion state at Hartree-Fock (HF) level is shown for illustrative purposes.   
\textbf{(b)}: All Siegert states of $^{2}A_{1}$ symmetry in the complex energy plane as a function of the N-H bond length. The points close to the real axis are unphysical pseudoresonances~\cite{ukrmol+}.
\textbf{(c)}: Magnitude of the contribution of the $\sigma^*$ orbital to the $^{2}A_{1}$ symmetry component of the scattering (Dyson-like) wavefunction for electron incoming in the direction of the N-H bond calculated as a function of the N-H bond distance and electron energy. The dashed orange and the dashed cyan line shows the contribution of the $\sigma^*$ orbital to the VBS and the DBS respectively.}
\label{fig:PES1}
\end{figure}

Both the resonances and virtual states can play role in DEA. Formally, these can be thought of as a generalization of bound state solutions with outgoing-wave boundary conditions $\approx\exp[\imath k.r]$ but with $k$ complex. This is the unified description of Siegert~\cite{siegert1939,tolstikhin_1998,batishchev2007,kukulin_1989} where bound, resonant and virtual states are treated on the same footing. In this formulation, bound states appear on the positive imaginary axis of momentum. Virtual and resonant states appear in the lower half complex plane of momentum and are therefore exponentially diverging solutions.

In this work, we have developed a method for locating and analyzing all Siegert states of the electron-molecule system in a unified way. The method, described in SI, is derived from the R-matrix approach for electron-molecule problems~\cite{ukrmol+,morgan_1988}. It uses a division of space to treat both exponentially decreasing (bound) and diverging (virtual, resonant state) distant parts of the Siegert state ($N+1$ electron) wavefunction analytically, while the short-range part is described using configuration-interaction techniques including bound and continuum orbitals. Both wavefunction parts are matched on the R-matrix sphere, which is large enough to fully contain the $N$-electron bound state of the target molecule.

The main results are shown in Figure~\ref{fig:PES1}.
In the complex plane plot (here showing only the states in the $^{2}A_{1}$ symmetry),  observable energy of a resonant state corresponds to its real part, its width is reflected in its imaginary part.
There are several near-threshold and resonant states in the system. First, the two $\pi^{*}$ resonances ($^{2}B_{1}$ at 2.32~eV and $^{2}A_{2}$ at 3.67~eV, in complex plane visible in SI, Fig.4). Second, there is a dipole-bound state (DBS) in $^{2}A_{1}$ symmetry. Third, there are two wide resonances of $\sigma^{*}$ character ($^{2}A_{1}$ and $^{2}B_{2}$) which correlate with the wide peaks in the C-H stretch vibrational excitation above 6~eV (Fig.~\ref{fig:2D}). The valence bound state (VBS) of $^{2}A_{1}$ symmetry appears for larger N-H bond lengths. The calculations also show the presence of a large number of very wide Siegert states which are an inherent property of quantum scattering~\cite{NUSSENZVEIG1959499}. 

The DEA in pyrrole can proceed in the $C_{2v}$ geometry without any geometry distortion via a process which we will call \textit{direct $\sigma^*$ mechanism}, where only the N-H bond is stretched. It is the same process which has been postulated to occur in many polar biomolecules~\cite{gallup_formic09,gallup_2013,fabrikant_uracil,Fabrikant_2015,scheer_2004,mcallister,Scheer_2005,oliveira10} and has been commonly denoted as being due to very broad dipole-supported $\sigma^*$ resonance. Its typical footprint are the sharp cusps in the vibrational excitation cross section such as observed in the N-H stretch excitation here.
We find the usage of the term resonance in this process to be a source of confusion. In the present case it would be represented by a repulsive $\sigma^{*}$(N-H) state as exemplified in Fig.~\ref{fig:PES1}a using the short dashed black line. However, no such state appears in Fig.~\ref{fig:PES1}b: trajectories of all resonances (states with Re(E)$>0$) are localized away from the real negative energy axis and none of them is connected with the final valence-bound state of the $^{2}A_{1}$ symmetry. Instead, the direct $\sigma^*$ mechanism results from the threshold divergence of the S-matrix~\cite{estrada_1984} which strongly influences the scattering continuum by a mechanism similar to a virtual state becoming bound~\cite{herzenberg1983}, see SI. When we evaluate the contribution of the $\sigma^*$ orbital to the $^2A_1$ symmetry component of the scattering (Dyson-like) wavefunction for electron incoming in the direction of the N-H bond (Fig.~\ref{fig:PES1}c, right axis), trajectory of its maximum (green line) resembles a repulsive potential curve. It can be used to parametrize DEA in Feshbach-type methods~\cite{domcke_1991} but it does not correspond to a physical dissociative resonant (Siegert) state.

\begin{figure*}[htb]
\includegraphics[width=1.0\textwidth]{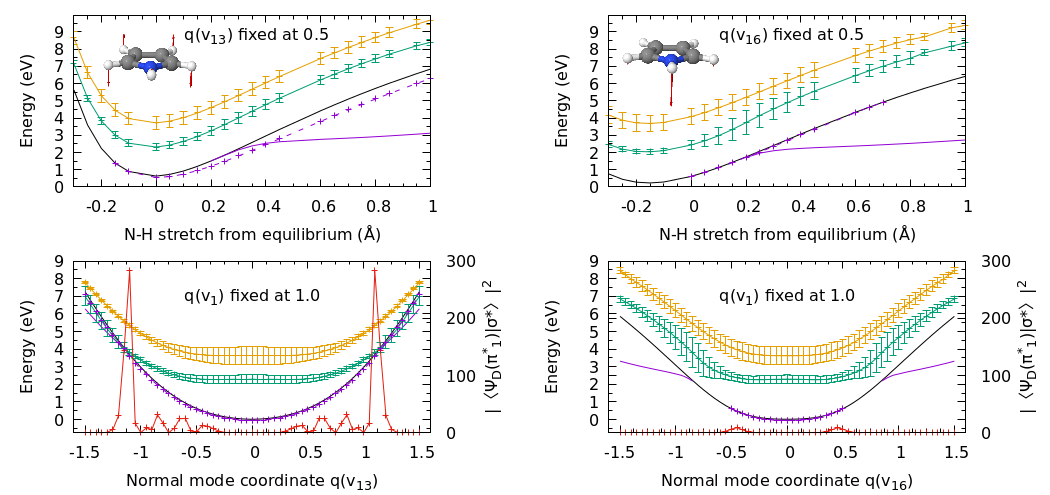}
\caption{The involved states (same as in Fig.~\ref{fig:PES1}) in the bent geometries. The modes $v_{13}$ (C-H bend), left, and $v_{16}$ (N-H bend) are depicted as insets. Top row: scans along the N-H stretch ($v_1$) in the geometries distorted to $q = 0.5$ in the two bending modes. Bottom row: scans along the two bending modes with the N-H distance fixed to the equilibrium distance, $q(v_{1})=1$. The red curve shows the magnitude of mixing of the $\pi^{*}_{1}$ resonant state with the lowest-lying $\sigma^{*}$ orbital that describes the valence bound state for large N-H bond lengths.}
\label{fig:PES2}
\end{figure*}

Even though the direct $\sigma^{*}$ DEA mechanism is open in pyrrole, it cannot explain the observed distant dissociation control which also involves other parts of the molecule.
We thus examine the behavior of the $\pi_1^*$ resonance upon breaking of the planar symmetry (Fig.~\ref{fig:PES2}), especially as a function of the two most important ring-distorting vibrational modes $v_{13}$ (C-H out-of-plane) and $v_{16}$ (N-H out-of-plane). The flat character of the $\pi_1^{*}$ resonance along these modes causes their efficient excitation. However, also in the broken symmetry this resonance remains non-dissociative in the N-H stretch direction, as in the C$_{2v}$ geometry. This is in contrast with the often assumed picture~\cite{oliveira10} described in the introduction, that bending opens the adiabatic dissociation pathway due to mixing of the $\pi^*$ resonance with the repulsive state.  

The involved states thus behave similarly as in the flat geometry. However, in the distorted geometries, the coupling between the resonance and the emerging valence bound state ($\sigma^{*}$ orbital) is allowed. Their mixing is represented by the red curve in Fig.~\ref{fig:PES2}. The motion along the $v_{13}$ mode is much more efficient than $v_{16}$ at coupling to the $\sigma^{*}$ mechanism (possibly because $v_{13}$ involves motion of four bonds, $v_{16}$ only of one). Additionally, the $\pi_1^{*}$ width behaves differently in the two bending directions.  For $v_{16}$, the  width increases in the region where the valence bound state forms. This leads to an enhanced detachment of electrons (this motion shakes the electrons off). The $v_{13}$ mode stabilizes the resonance (reduces its width and brings it closer in energy to the neutral state). It is reasonable to conclude that these factors combine to make $v_{13}$ motion more efficient at enhancing DEA cross section than the other modes. That makes the C-H motion the limiting step in the dynamics.

Finally, as we show in SI, DEA in pyrrole is strongly directionaly sensitive due to the different symmetries of the $\sigma^*$ and $\pi^*$ orbitals involved: the direct $\sigma^*$ mechanism is maximized for electrons incoming along the NH bond while the distant symmetry breaking mechanism proceeding via the $\pi^*$ orbital requires electrons incoming from directions not in the molecular plane. This observation paves the way to disentangling both contributions and shows that the role of the $\pi^{*}$ resonance is to temporarily capture electrons incoming from many different directions in order to give them enough time to ``discover" the non-resonant $\sigma^{*}$ pathway. This discovery time is limited by the resonance half-time, which in this case is approximately 5.5~fs. Therefore the nuclear motion which is key for enabling DEA in pyrrole must occur on an ultrafast timescale of only a few femtoseconds.

In conclusion, we have demonstrated that the N-H bond cleavage in pyrrole induced by low energy electrons is controlled by motion of the C-H bonds which are not being broken. The Siegert-state analysis provides interpretation of this effect. 
The current findings:
\begin{enumerate}[(i)]
\item Resolve a long-standing discussion about the role of the $\sigma^*$ mechanism in the DEA to polar biomolecules. While the postulation and parametrization of broad $\sigma^*$ resonances has succesfully reproduced DEA cross sections in many molecules,~\cite{gallup_formic09, vizcaino12, fabrikant_uracil, Fabrikant_2015} the scattering calculations consistently fail to locate such resonances. The hypothesis has been that these are too far from the real axis (they are too broad) to be discerned. We show that in pyrrole the  $\sigma^*$ mechanism is open but it is non-resonant and is connected with the presence of virtual state.  
\item Raise questions regarding DEA in biological systems (it is the only process in radiation damage which can cause bond breaking at energies lower than those of electronic excitation) and the role of environment in inhibiting or amplifying the symmetry-breaking motion of peripheral parts of the molecule.
\item Show the need for multi-dimensional and multi-state models of resonant nuclear dynamics in polyatomic molecules which will be able to rationalize and predict this type of motion.
\end{enumerate}

This work is part of the Czech Science Foundation projects 20-11460S (J.F.) and 20-15548Y (Z.M.)  Z. M. also acknowledges support from Charles University (PRIMUS/20/SCI/003), OP RDE project CZ.02.2.69/0.0/0.0/16$\_$027/0008495, the MEYS projects "e-Infrastructure CZ – LM2018140" and "IT4Innovations National Supercomputing Center - LM2015070". We thank R. \v{C}ur\'{i}k, K. Houfek and M. \v{C}\'{i}\v{z}ek for stimulating discussions.


\end{document}